\RequirePackage{fix-cm}
\documentclass[smallextended, final]{svjour3}          
\smartqed  

\usepackage{paralist, enumerate, enumitem}
\usepackage{amssymb, amsmath, bm}
\usepackage{geometry}
\usepackage{graphicx, subfigure}
\usepackage{booktabs, multirow}
\usepackage{cite}

\usepackage{color}

\makeatletter
\newcommand{\dotminus}{\mathbin{\text{\@dotminus}}}

\newcommand{\@dotminus}{%
  \ooalign{\hidewidth\raise1ex\hbox{.}\hidewidth\cr$\m@th-$\cr}%
}
\makeatother
\graphicspath{{figures/}}
\newlength\imagewidth
\newtheorem{Fact}{Fact}

\newcommand\SDM{{\rm SDM}}
\newcommand\sort{{\rm sort}}

\DeclareMathOperator\f{f}

\journalname{Multimedia Tools and Applications}

\begin{document}

\title{Cryptanalyzing an image {encryption} algorithm based on {scrambling} and Vegin{\`e}re cipher}

\author{Li Zeng         \and
        Renren Liu      \and
        Leo Yu Zhang    \and
        Yuansheng Liu   \and
        Kwok-Wo Wong
}

\institute{Li Zeng, Renren Liu, Yuansheng Liu \at
              College of Information Engineering, Xiangtan University, Xiangtan 411105, Hunan, China \\
              \email{lily173864258@gmail.com}            \\
\\
            Leo Yu Zhang,  Kwok-Wo Wong \at
            Department of Electronic Engineering, City University of Hong Kong, Hong Kong \\
}

\date{Received: date / Accepted: date}

\maketitle

\begin{abstract}
Recently, an image encryption algorithm based on scrambling and Vegin{\`e}re cipher has been proposed.
However, it was soon cryptanalyzed by Zhang \textit{et al.} using a combination of chosen-plaintext attack and differential attack.
This paper briefly reviews the two attack methods proposed by Zhang \textit{et al.} and outlines the mathematical interpretations of them.
Based on their work, we present an improved chosen-plaintext attack to further reduce the number of chosen-plaintexts required, which is proved to be optimal.
Moreover, it is found that an elaborately designed known-plaintex attack can efficiently compromise the image cipher under study.
This finding is verified by both mathematical analysis and numerical simulations. The cryptanalyzing techniques described in this paper may
provide some insights for designing secure and efficient multimedia ciphers.

\end{abstract}

\section{Introduction}
\label{sec:intro}


{The rapid development of computer networks enables us to enjoy multimedia contents such as image and video conveniently.
However, it also leads to challenges to the security of multimedia data which are transmitted over public channels.}
Due to bulk data volume and high correlation among neighboring pixels/frames of the raw image/video data,
traditional encryption techniques, such as AES, 3DES and IDEA, are not appropriate for {image/video} encryption.
The improperness appears in the following scenarios:
\begin{itemize}
\item The block size of traditional block ciphers is too small when comparing to the amount of multimedia data to be
encrypted. For natural gray-scale images at a resolution of $1024\times 1024$, it requires the execution of 3DES for more than $10^5$ times to encrypt
one single image. The efficiency problem makes traditional block ciphers inappropriate for real-time applications, such as online TV,
video conferencing, etc.
\item Generally, the security level of traditional block ciphers is higher than that required in multimedia data encryption.
For the protection of commercial movies, it merely requests that breaking the cipher will cost the attacker more than that for buying one genuine copy of the movie.
In such scenario, some lightweight encryption algorithms, such as perceptual encryption \cite{li2007design} and selective encryption \cite{kim2011selective}, are
competent for this purpose.
\item The strong correlation among adjacent pixels/frames of image/video cannot be thoroughly removed using traditional block ciphers in some operation modes.
We give an example to illustrate this phenomenon. The cartoon image shown in Fig.~\ref{fig:qe:plainimage} is encrypted using DES under the electronic codebook mode,
the corresponding cipher-image is depicted in Fig.~\ref{fig:qe:cipherimage}.
It is clear that the shape of the cartoon image can be recognized from the cipher-image directly without decryption.

\begin{figure}[!htb]
\centering
\subfigure[]{
    \label{fig:qe:plainimage}
    \begin{minipage}[t]{\imagewidth}
    \centering
    \includegraphics[width=\imagewidth]{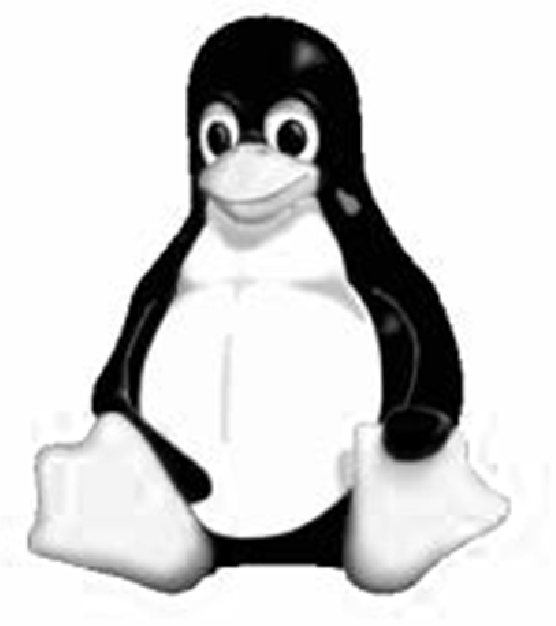}
    \end{minipage}}
\subfigure[]{
    \label{fig:qe:cipherimage}
    \begin{minipage}[t]{\imagewidth}
    \centering
    \includegraphics[width=\imagewidth]{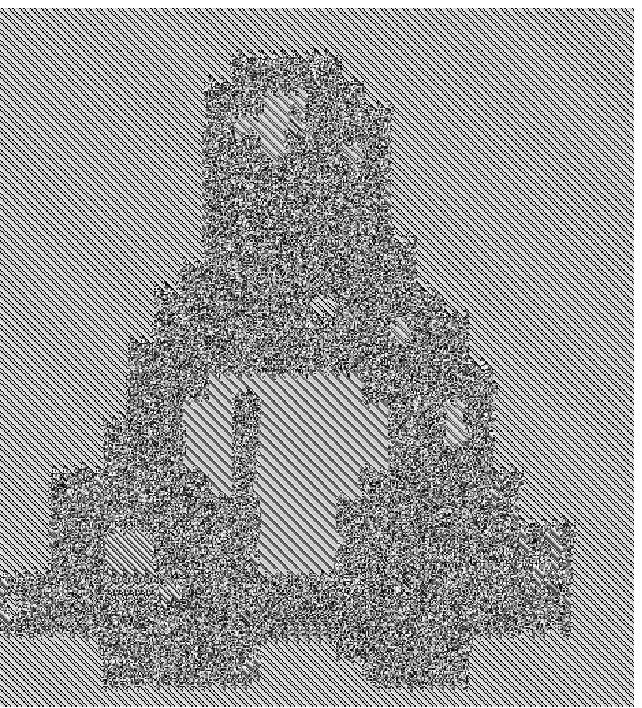}
    \end{minipage}}
\caption{
(a) A cartoon image;
(b) cipher-image of Fig.~\ref{fig:qe:plainimage} using DES under the electronic codebook mode.}
\label{fig:qe}
\end{figure}

\end{itemize}

Chaos, which was extensively studied since 1960s, appears to be a promising solution to the above mentioned challenges as
some intrinsic characteristics of chaotic maps, such as sensitivity and ergodicity,
coincide with the confusion and diffusion properties of a good cryptographic algorithm \cite{Shannon:Communication:Bell49}.
Consequently, many chaos-based encryption algorithms \cite{fridrich1998symmetric, chen2004symmetric, jakimoski2001chaos,
mao2004novel, behnia2008novel, riad2012new} have been proposed in the past decade.
At the same time, the cryptanalyses of these ciphers also received considerable research attention
\cite{alvarez2004cryptanalysis, chen2006chosen, solak2010cryptanalysis, li2012breaking, li2009improving, zhangcryptanalyzing}.
When a chaotic system is implemented using finite precision computation, it suffers seriously from the so-called dynamical degradation,
which accounts for the phenomenon that some dynamical properties are substantially different from those found in the continuous setting \cite{li2005dynamical}.
A typical cryptanalysis work based on the dynamical degradation of chaotic functions was presented in \cite{li2003security}.



Aims to bypass the intractable dynamical degradation problem, Li \textit{et al.} proposed a novel image encryption algorithm based on a
$2$D coupled logistic map \cite{li2012image}. Instead of using quantized output sequences of the employed chaotic map, which is a common method employed by
most chaotic ciphers, two random sequences are generated by means of sorting the chaotic outputs. Then, one of the random sequences is
used to mask the plain-image as performed in the Vegin{\`e}re cipher and the remaining one is used to further scramble the previous output.
Intuitively, this cipher is not as secure as the authors claimed in \cite[Sec.~3]{li2012image} since it does not possess sufficient
avalanche effect \cite{Wade:IntroCypt:Prentice2002}.

In \cite{zhang2014cryptanalysis}, Zhang \textit{et al.} suggested two attacks to compromise the scheme in \cite{li2012image}.
They can be considered as a combination of chosen-plaintext attack and differential attack.
Though their attacks are feasible in theory, they require a large number of
chosen plain-images, and so the computation complexity is high. By breaking the equivalent key streams in reverse order as suggested by Zhang
\textit{et al.}, we propose a chosen-plaintext attack which is optimal in terms of the required number of chosen plain-images.
Moreover, we present an elaborately designed known-plaintext attack to break this encryption algorithm efficiently.

The rest of this paper is organized as follows. The next section
introduces the original image encryption algorithm briefly.
In Section \ref{sec:cryptanalysis}, two attack methods proposed
by Zhang \textit{et al.} are reviewed. Then we present the proposed optimal
chosen-plaintext attack and the efficient known-plaintext attack both theoretically and numerically.
Some concluding remarks are drawn in the last section.

\section{The original image encryption algorithm}
\label{sec:alogrithm}
In this section, we describe the image cipher of \cite{li2012image} in a concise way with
the criterion that  its security is not changed. Some simulation results are presented after the description.
Given the secret key $(x_0, y_0,\mu_1, \mu_2, \gamma_1, \gamma_2)$,
the cipher operates as follows:

\begin{enumerate}
\item Input an $8$-bit gray-scale image of size $L$ and convert it to a
one-dimensional sequence $P=\{p(i)\}_{i=1}^{L}$ in raster scan order.

\item Generate two sequences $\{x_i\}_{i=1}^{L}$ and $\{y_i\}_{i=1}^{L}$
using the $2$D coupled logistic map given by Eq.~\eqref{eq:2dlm}
with initial condition $\{x_0, y_0\}$ and the control parameter
$\{\mu_1, \mu_2, \gamma_1, \gamma_2\}$,
\begin{equation}
\label{eq:2dlm}
\begin{cases}
x_{i+1} = \mu_1x_i(1-x_i)+\gamma_1y_i^2,\\
y_{i+1} = \mu_2y_i(1-y_i)+\gamma_2(x_i^2+x_iy_i).
\end{cases}
\end{equation}

\item Sort the two sequences $\{x_i\}_{i=1}^{L}$ and $\{y_i\}_{i=1}^{L}$
to obtain
\begin{equation}
\nonumber
\begin{split}
[U, \hat{X}] = \sort(\{x_i\}_{i=1}^{L}),\\
[V, \hat{Y}] = \sort(\{y_i\}_{i=1}^{L}),
\end{split}
\end{equation}
where $\hat{X}$ and $\hat{Y}$ are the resultant sequences after
sorting $\{x_i\}_{i=1}^{L}$ and $\{y_i\}_{i=1}^{L}$ in ascending order, respectively,
$U=\{u(i)\}_{i=1}^{L}$ and $V=\{v(i)\}_{i=1}^{L}$ are their corresponding index values.

\item Compute the corresponding pixel value of the cipher-image according to the following formula:
\begin{equation}
\label{eq:encrypt}
c(v(i)) = p(i)\dotplus u(i),
\end{equation}
where $i \in \{1, 2, \cdots, L\}$ and $(a \dotplus b) = (a+b)\mod{256}$.

\item Rearrange the one-dimensional sequence $\{c(i)\}_{i=1}^{L}$ to a two-dimensional matrix
row by row and the cipher-image is obtained.
\end{enumerate}

\begin{figure}[!htb]
\centering
\subfigure[]{
    \label{fig:Baboon}
    \begin{minipage}[t]{\imagewidth}
    \centering
    \includegraphics[width=\imagewidth]{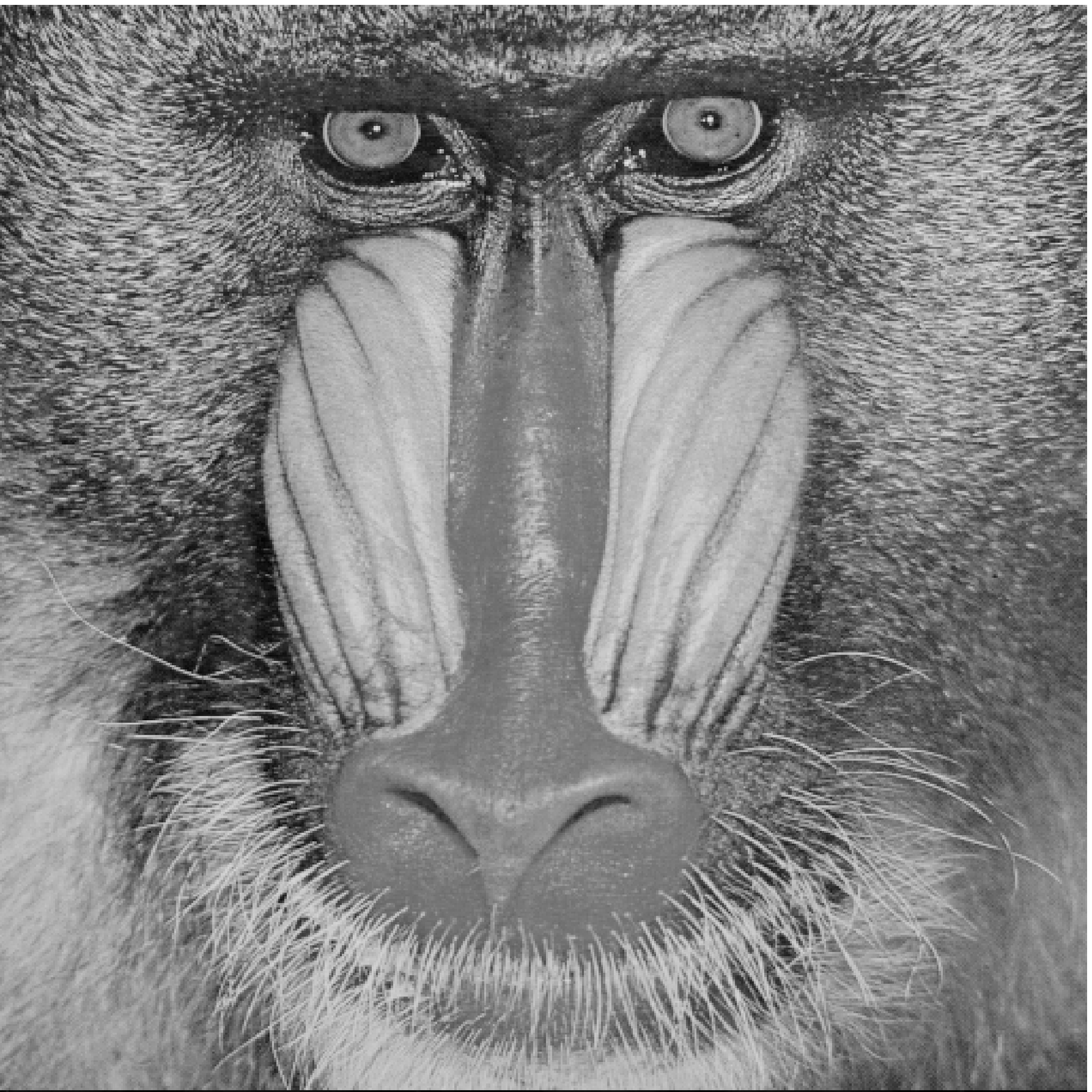}
    \end{minipage}}
\subfigure[]{
    \label{fig:cBaboon}
    \begin{minipage}[t]{\imagewidth}
    \centering
    \includegraphics[width=\imagewidth]{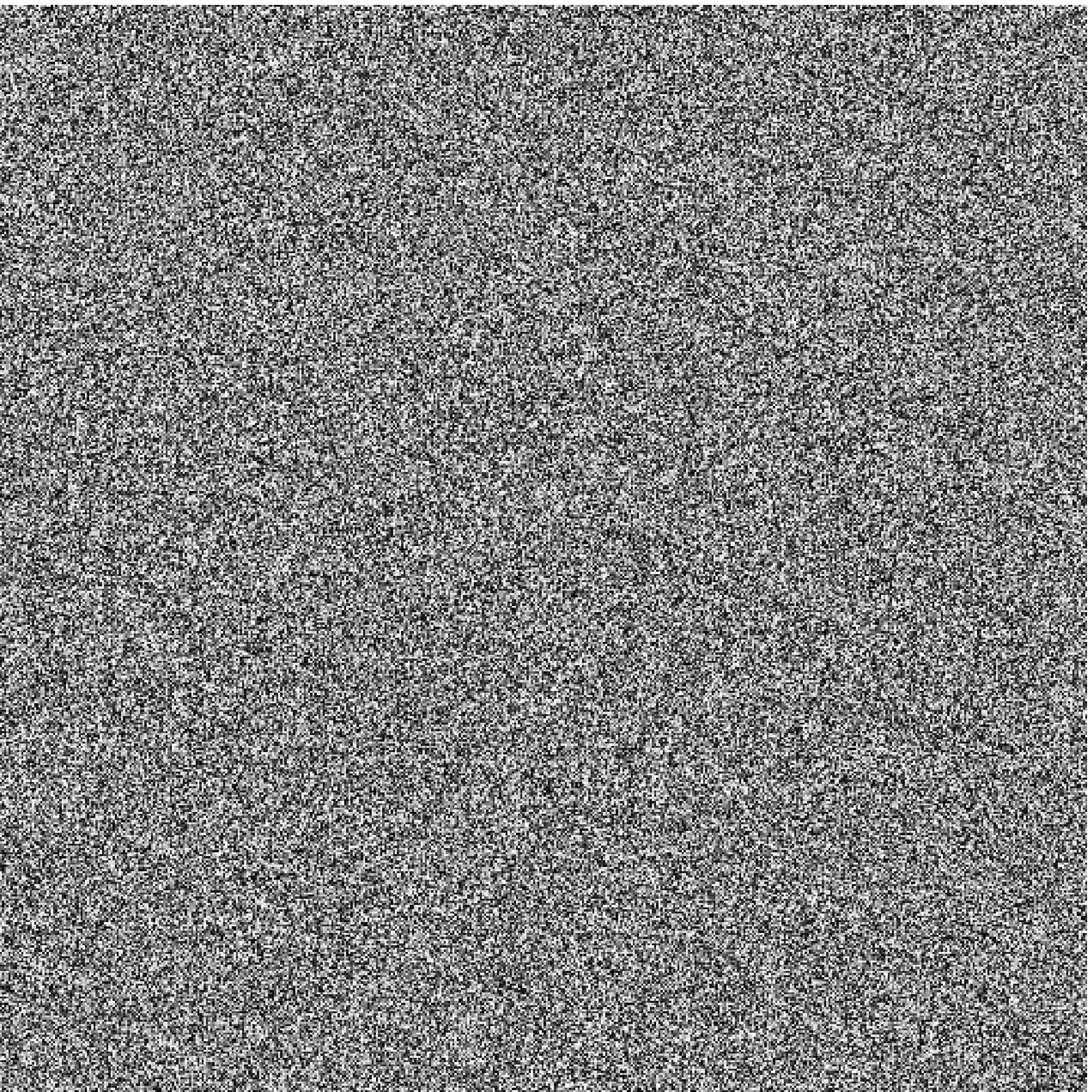}
    \end{minipage}}
\subfigure[]{
    \label{fig:Lenna}
    \begin{minipage}[t]{\imagewidth}
    \centering
    \includegraphics[width=\imagewidth]{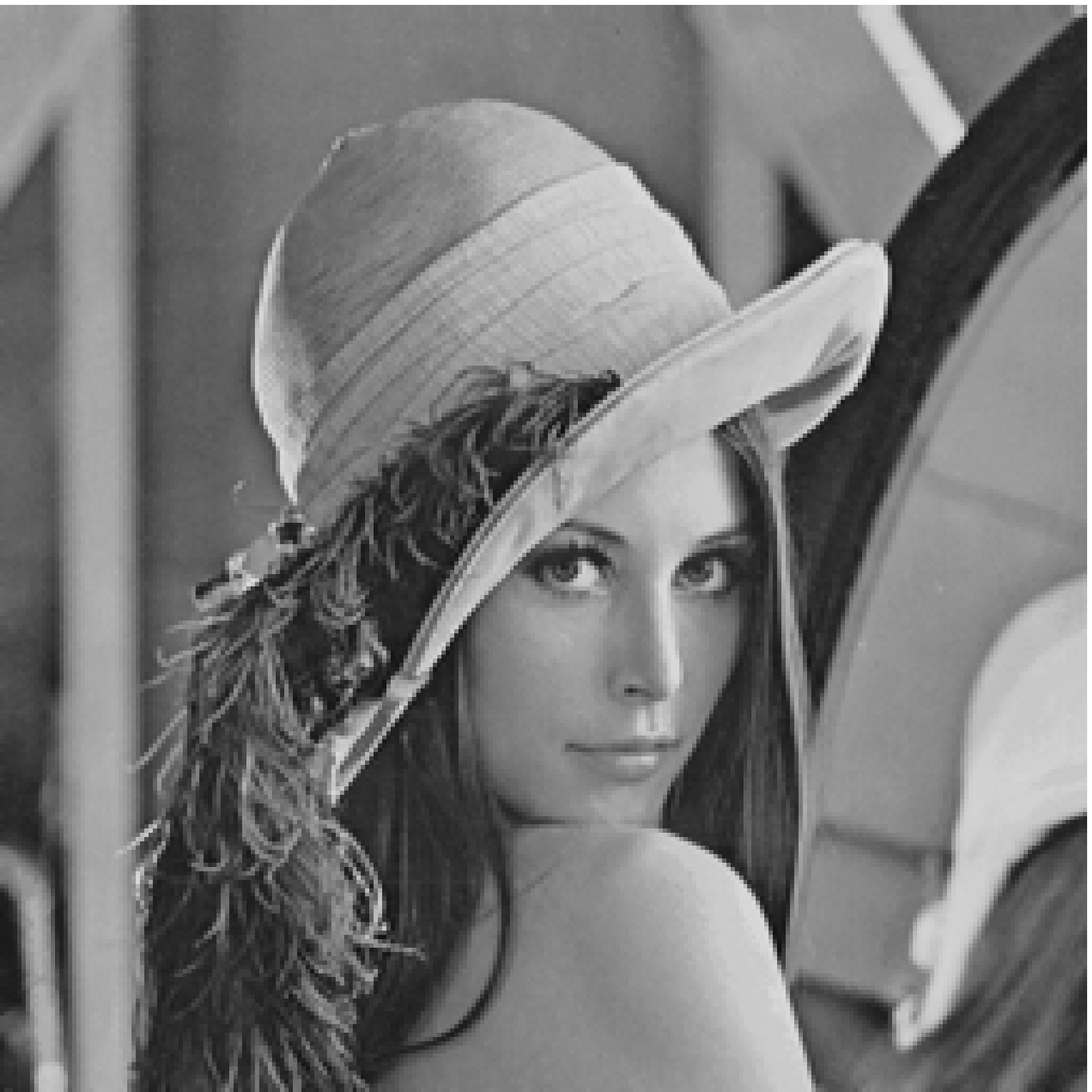}
    \end{minipage}}
\subfigure[]{
    \label{fig:cLenna}
    \begin{minipage}[t]{\imagewidth}
    \centering
    \includegraphics[width=\imagewidth]{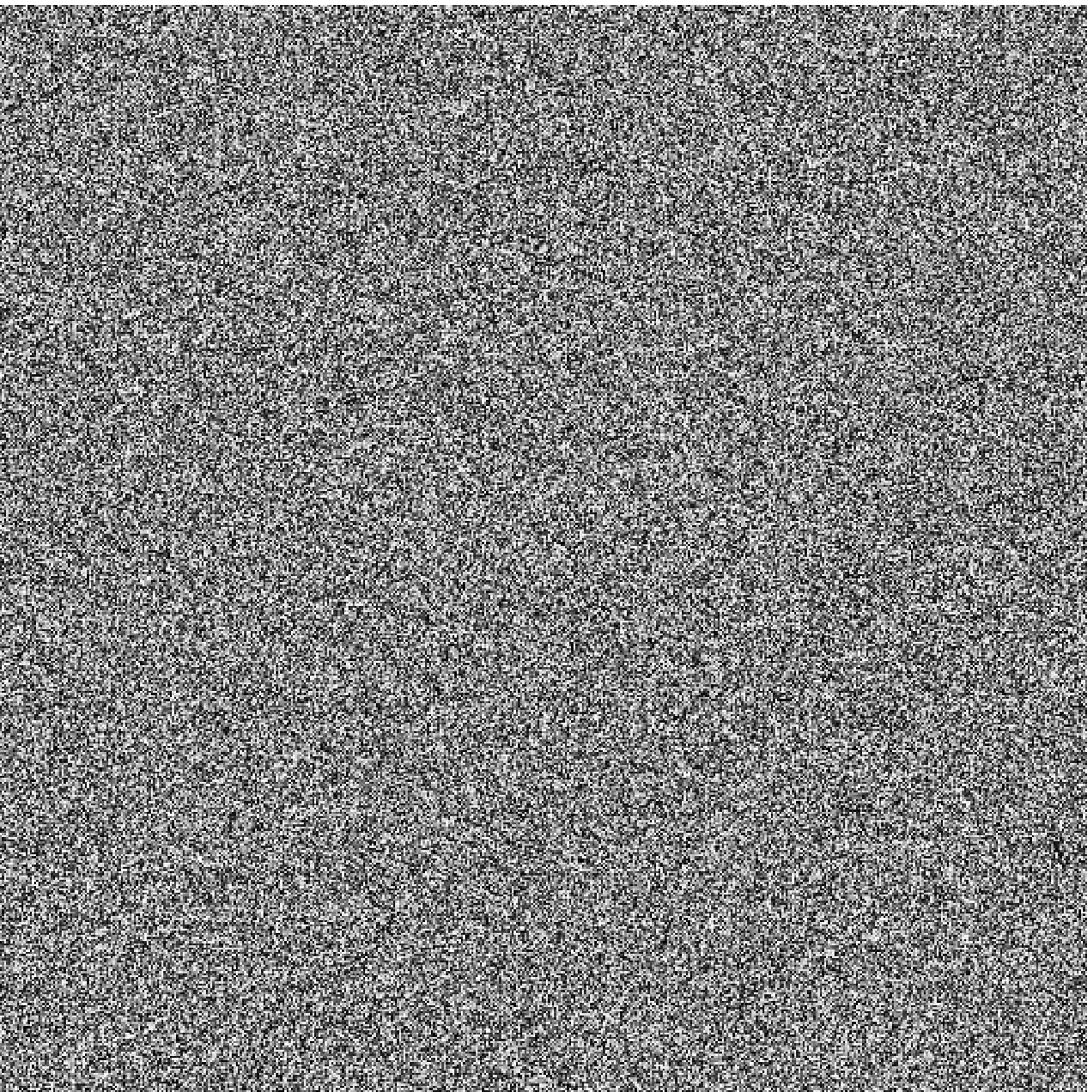}
    \end{minipage}}
\caption{Two plain-images and their corresponding cipher-images:
(a) plain-image ``Baboon";
(b) cipher-image of ``Baboon";
(c) plain-image ``Lenna";
(d) cipher-image of ``Lenna".}
\label{fig:encryption}
\end{figure}

We are not going to describe the detailed decryption algorithm since it is very similar to its encryption counterpart.
Two $512 \times 512$ plain-images, ``Baboon" and ``Lenna" depicted in Fig.~\ref{fig:Baboon} and Fig.~\ref{fig:Lenna}, respectively,  are encrypted
using the secret key $(x_0, y_0, \mu_1, \mu_2, \gamma_1, \gamma_2) = (0.02145, 0.3678, 2.93, 3.17, 0.179, 0.139)$,
which is identical to the key chosen in \cite[Sec.~4.1]{li2012image}.
The cipher-images are shown in Fig.~\ref{fig:cBaboon} and Fig.~\ref{fig:cLenna}, respectively.

\section{Cryptanalysis}
\label{sec:cryptanalysis}
In the original paper \cite{li2012image}, the authors claimed that
the initial condition $\{x_0, y_0\}$ and the control parameters
$\{\mu_1, \mu_2, \gamma_1, \gamma_2\}$ of the $2D$ coupled logistic
map should serve as the secret key to guarantee a huge key space
to resist brute-force attacks. From the cryptanalytic point of view, our objective
is to reveal the equivalent encryption keystreams $\{u(i)\}_{i=1}^{L}$ and $\{v(i)\}_{i=1}^{L}$ \cite[Sec.~3]{zhang2014cryptanalysis},
rather than finding the exact initial key $(x_0, y_0,\mu_1, \mu_2, \gamma_1, \gamma_2)$.
Also, is is commonly believed that iterating a chaotic system reversely from its output is computational intractable.

Obviously, the two sequences $\{u(i)\}_{i=1}^{L}$ and $\{v(i)\}_{i=1}^{L}$ are
identical to the secret key when the algorithm is used to
encrypt plain-images of the same size. According to $\mathbf{Fact}$~\ref{fact} and
the encryption fromula~\eqref{eq:encrypt}, we know that the sequence
$\{u(i)\}_{i=1}^{L}$ is equivalent to the sequence $\{k(i)\}_{i=1}^{L}$
in the encryption process if $k(i) = u(i)\mod{256}$.
Now, we can rewrite the encryption equation (\ref{eq:encrypt}) as
\begin{equation}
\label{eq:ReEncrypt}
c(v(i)) = p(i)\dotplus k(i).
\end{equation}
\begin{Fact}
\label{fact}
$(a\dotplus b) = (a \dotplus (b\mod{256})).$
\end{Fact}

Taking these factors into consideration, we are now able to compromise the cipher under study.
Section~\ref{subsec:yushuidea} presents some cryptanalysis work performed by Zhang \textit{et al} \cite{zhang2014cryptanalysis}. We briefly review their attacks and
provide the mathematical interpretations of \textit{Method II} in \cite[Sec.~3.2]{zhang2014cryptanalysis} based on a simple
fact\footnote{Instead of proving the effectiveness of \textit{Method II} mathematically, the authors of \cite{zhang2014cryptanalysis} solved the problem by trying all possible combinations.}.
Section.~\ref{subsec:cpa} presents an optimal chosen plaintext attack using the minimum number of chosen plain-images.
This is a direct application of the result reported in \cite{li2008general, li2011optimal}.
In Sec.~\ref{subsec:kpa}, we focus on the cryptanalysis of this cipher under a known plain-image attack scenario.
Theoretical analyses and experimental results are provided to demonstrate the effectiveness of our attacks.

\subsection{Attacks proposed by Zhang \textit{et al.} }
\label{subsec:yushuidea}
The chosen-plaintext attack is a fundamental attack scenario which plays a significant role in evaluating the security of a cipher.
In this attack scenario, the attackers have the freedom to choose any plaintexts
to be encrypted and obtain the corresponding ciphertexts. Differential attack, which was firstly proposed by Biham and Shamir in \cite{Biham:Deslike:Crypt90} for cracking DES, is an
effective tool to analyze a cipher with Feistel structure. It is also found useful for analyzing other encryption algorithms \cite{li2012breaking,zhang2012cryptanalyzing}.

In \cite{zhang2014cryptanalysis}, Zhang \textit{et al.} suggested two methods to break the cipher under study using a combination of
chosen plain-image attack and differential attack. The basic ideas behind these methods are the same but the second method requires fewer
chosen plain-images.


The first method operates as follows.
Choose a dark image $P=\{p(i)\}_{i=1}^ {L}$ whose pixel values are all zero.
Then choose another plain-image $P'=\{p'(i)\}_{i=1}^ {L}$ with only one pixel different from $P$,
e.g., $p'(1)=1$ and $p'(i) \equiv 0$ for all $i>1$.
Encrypt these two images and denote the corresponding cipher-images as $C=\{c(i)\}_{i=1}^ {L}$ and $C'=\{c'(i)\}_{i=1}^ {L}$, respectively.
According to the encryption
formula given by Eq.~\eqref{eq:ReEncrypt}, it can be  concluded that,
only one pair of pixel elements are different in the two corresponding
cipher-images. Making use of differential relationship of the cipher-image pixels,
we can formulate this process as
\begin{eqnarray}
c(v(1)) \dotminus c'(v(1)) & =  (0 \dotplus k(1)) \dotminus (1 \dotplus k(1)) & \neq  0, \nonumber\\
c(v(i)) \dotminus c'(v(i)) & =  (0 \dotplus k(i)) \dotminus (0 \dotplus k(i)) & = 0, \nonumber
\end{eqnarray}
where $i>1$ and $(a \dotminus b)=(a-b+256) \mod 256$. Thus, it is easy to identify $v(1)$ by finding the nonzero element
of the difference image between $C$ and $C'$. Moreover, one can determine $k(1)$
by $k(1)= c(v(1)) \dotminus p(1)$.
Repeat the process for $(L-1)$ more times using different chosen plain-images who have
only one pixel different from the dark image, one can finally reveal all the
equivalent key streams $\{k_i\}_{i=1}^{L}$ and $\{v_i\}_{i=1}^{L}$ at the
cost of $(1+L)$ chosen-plain images.

The second method improves the first one in terms of the number of chosen-plain
images based on the fact that a gray-scale image has $256$ different pixel values.
Randomly set $255$ pixels different from $P$ having gray values $\{1,2,\cdots, 255\}$, and denote this chosen-image as $P'$.
Referring to $\mathbf{Fact}$~\ref{fact2}, it is easy to conclude that the difference between $P$ and $P'$ is exactly the same
as the difference of their cipher-images, but the locations are shuffled by the key stream  $\{v(i)\}_{i=1}^{L}$.
According to the bijection relationship of the $255$ different gray values between difference of plain-images and difference of cipher-images,
one can obtain $255$ distinct position relationships and thus the corresponding values of $k(i)$.
Therefore, the image scrambling algorithm can be broken with
$(1 + \lceil L/255\rceil)$ chosen-plain images.
\begin{Fact}
\label{fact2}
$\f(x) = (x \dotplus k) \dotminus k = x$, where $k$ and $x$ are integers in the interval $[0, 255]$.
\end{Fact}

\begin{figure}[!htb]
\centering
\subfigure[]{
    \label{fig:chosendark}
    \begin{minipage}[t]{\imagewidth}
    \centering
    \includegraphics[width=\imagewidth]{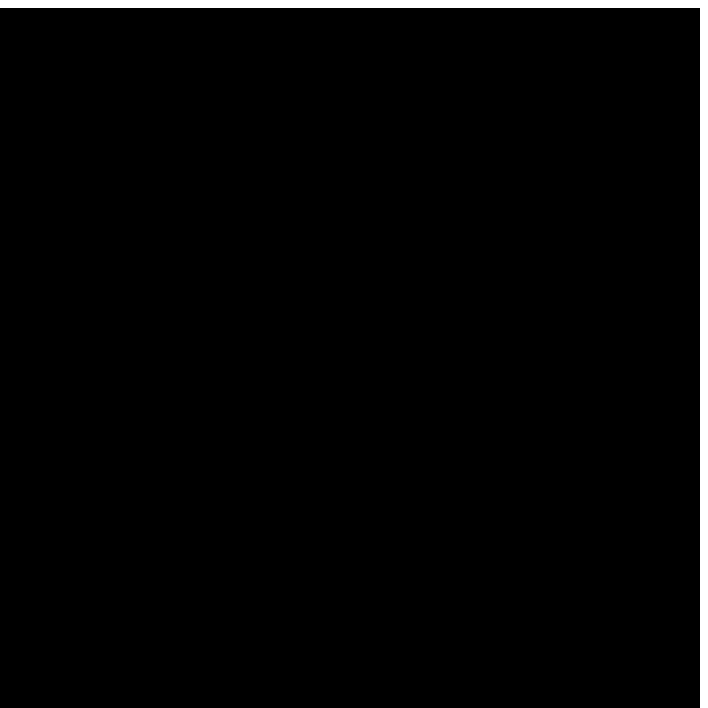}
    \end{minipage}}
\subfigure[]{
    \label{fig:chosen0-255}
    \begin{minipage}[t]{\imagewidth}
    \centering
    \includegraphics[width=\imagewidth]{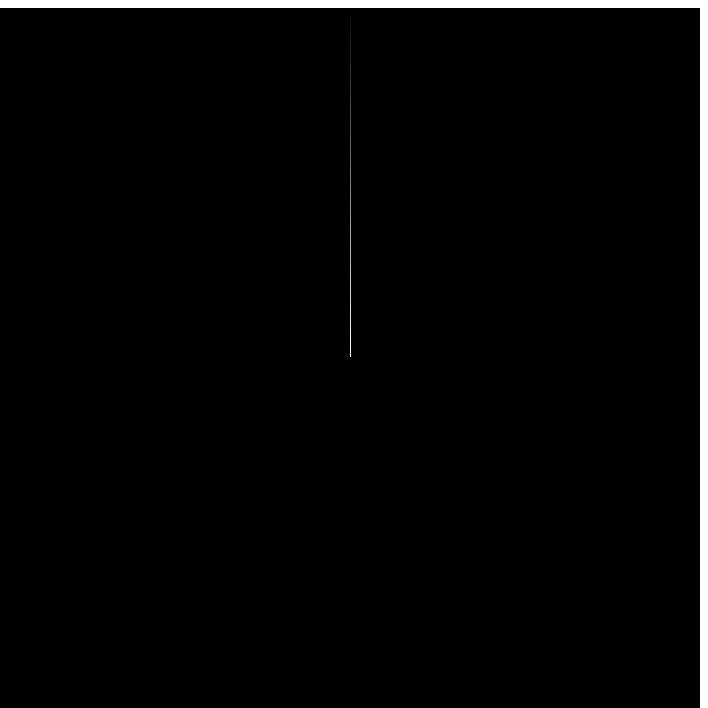}
    \end{minipage}}
\subfigure[]{
    \label{fig:cipherdifference}
    \begin{minipage}[t]{\imagewidth}
    \centering
    \includegraphics[width=\imagewidth]{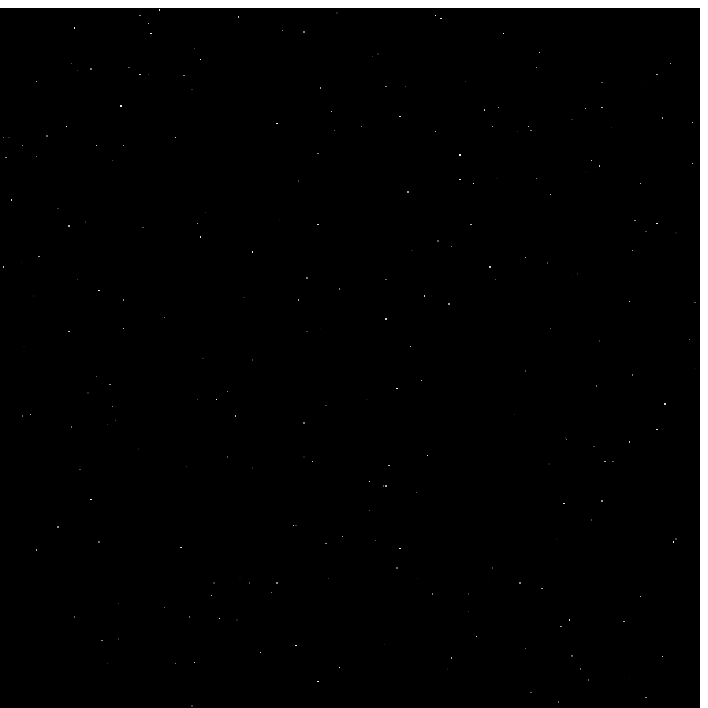}
    \end{minipage}}
\subfigure[]{
    \label{fig:recoverBaboon}
    \begin{minipage}[t]{\imagewidth}
    \centering
    \includegraphics[width=\imagewidth]{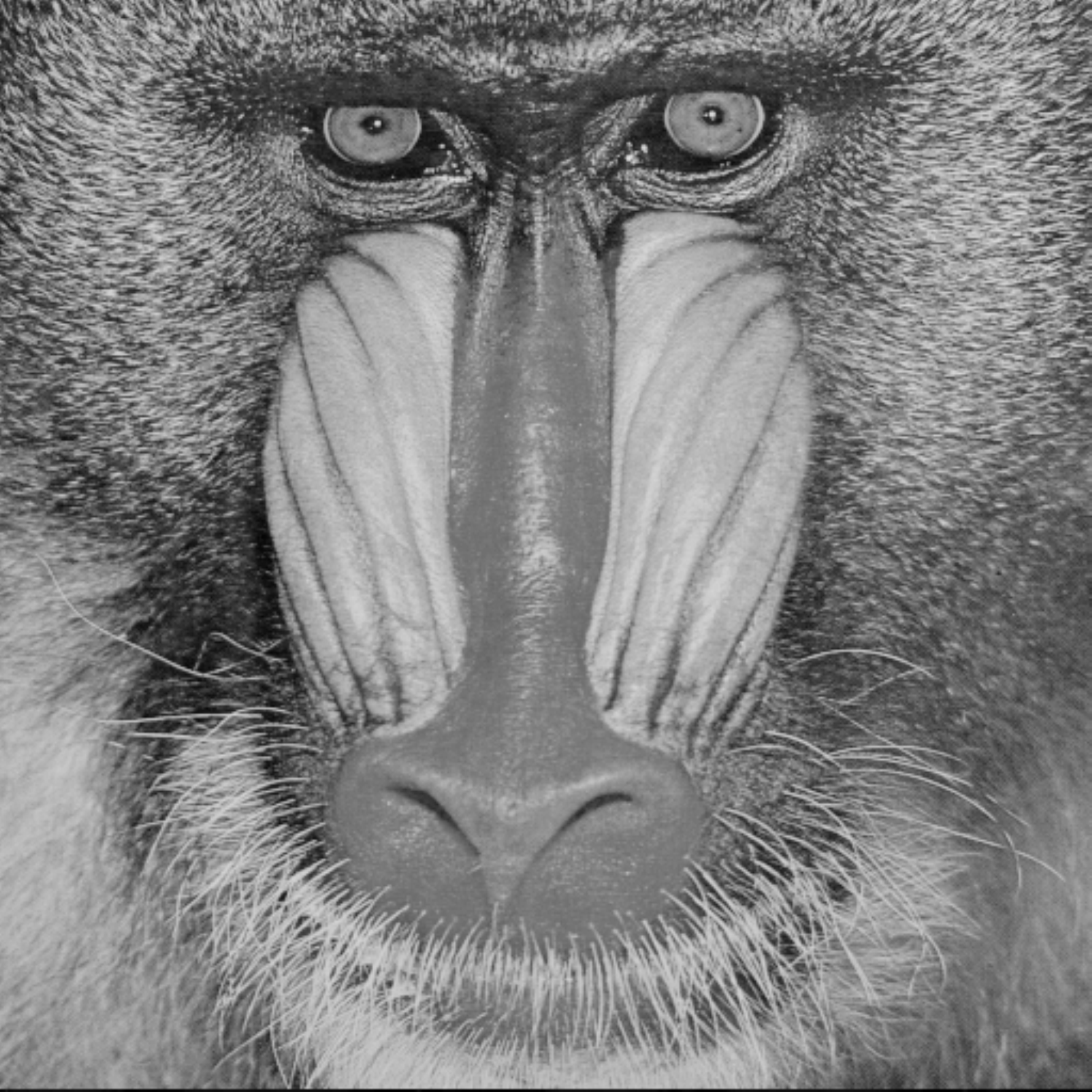}
    \end{minipage}}
\caption{Test of \textit{Method II} in \cite{zhang2014cryptanalysis}:
(a) the dark chosen plain-image;
(b) a modified chosen plain-image;
(c) the difference between cipher-images of Fig~\ref{fig:chosendark} and Fig~\ref{fig:chosen0-255};
(d) the image recovered from the cipher-image shown in Fig.~\ref{fig:cBaboon}.}
\label{fig:yushuattack2}
\end{figure}

The two chosen plain images shown in Figs.~\ref{fig:chosendark} and \ref{fig:chosen0-255} are encrypted using the key selected in \cite[Sec.~4.1]{li2012image}.
The difference of the two cipher-images\footnote{Perceptually, Fig.~\ref{fig:cipherdifference} is identical to Fig.~\ref{fig:chosendark}. But
there are $255$ nonzero pixels uniformly distributed in Fig.~\ref{fig:cipherdifference} while  Fig.~\ref{fig:chosendark} does not.},
which is shown in Fig.~\ref{fig:cipherdifference}, are used to recover $255$ unknowns of the key stream $\{v(i)\}_{i=1}^{L}$.
Repeat this test for $(\lceil L/255\rceil-1)$ more times using other chosen plain-images, the equivalent key streams used for encryption can be revealed completely.
The recovered key streams are further used to attack the cipher-image depicted in Fig.~\ref{fig:cBaboon} and the result is shown in Fig.~\ref{fig:recoverBaboon}.
The retrieved image is exactly the same as the original image ``Baboon".

\subsection{Optimal chosen-plaintext attack}
\label{subsec:cpa}
As described in Sec.~\ref{subsec:yushuidea}, the attacks suggested in \cite{zhang2014cryptanalysis} retrieve the equivalent secret key $v(i)$ and $k(i)$
sequentially, i.e., recover $v(i)$ first and then $k(i)$. Here, we suggest recovering $v(i)$ and $k(i)$ in a reversed order. In this way, the optimality
of the chosen plain-image attack is achieved.

Without loss of generality, suppose that there exists a random sequence $\{r(j)\}_{j=1}^{L}$ such that
\begin{equation*}
r(v(i)) = k(i),
\end{equation*}
where $\{v(i)\}_{i=1}^{L}$ is the undetermined equivalent key stream. Substitute $r(j)$ into Eq.~\eqref{eq:ReEncrypt},
we have
\begin{equation}
\label{eq:remarked}
c(v(i)) = p(i) \dotplus r(v(i)).
\end{equation}
First, choose a plain-image $P$ with constant pixel values, i.e., $P = \{p(i)\equiv d\}_{i=1}^{L}$ and $d\in [0, 255]$. Then
get the corresponding cipher image $C=\{c_i\}_{i=1}^{L}$.
Referring to Eq.~\eqref{eq:remarked}, we can obtain the equivalent key stream $\{r(j)\}_{j=1}^{L}$ by solving
\begin{equation}
\nonumber
r(i) = d \dotminus c(i),
\end{equation}
where $i =1,2,\cdots, L$.

Once the sequence $\{r(i)\}_{i=1}^{L}$ has been recovered, the image encryption algorithm under study degrades to a
permutation-only encryption algorithm. Referring to the cryptanalysis
of permutation-only encryption algorithms \cite{li2008general, li2011optimal},
$\lceil (\log_2L)/8\rceil$ pairs of chosen plain-images are sufficient to recover the rest equivalent key sequence $\{v(i)\}_{i=1}^{L}$.
The optimality of the proposed chosen plain-image attack is straightforward since we only require one chosen image to recover $\{r(i)\}_{i=1}^{L}$
and its optimality on permutation-only cipher has already been proven in \cite{li2011optimal}.

\subsection{The proposed known-plaintext attack}
\label{subsec:kpa}
In a known-plaintext attack, the attacker possesses some samples of both the plaintext and the corresponding ciphertext.
Different from the chosen-plaintext attack, the attacker is not allowed to choose the plaintext to be encrypted.
In other words, if the attacker inputs a message with elaborately designated structures for encryption, a trusted third party or the encryption machine will decline this request.
Generally speaking, cryptanalysis based on known-plaintext attack is more difficult than that using chosen-plaintext attack.

Assume that two plain-images $P_1=\{p_1(i)\}_{i=1}^{L}, P_2=\{p_2(i)\}_{i=1}^{L}$
and the corresponding cipher-images $C_1=\{c_1(i)\}_{i=1}^{L}, C_2=\{c_2(i)\}_{i=1}^{L}$
encrypted with the same secret key are available.
Obviously, for any $i, j\in [1, L]$, if $\Delta_p \doteq (p_1(i)\dotminus p_2(i)) = \Delta_c \doteq (c_1(j)\dotminus c_2(j))$,
one can realize that $j$ is a possible solution of $v(i)$.
As $\Delta_c \in [0, 255] \ll L$ {and the pixel values of the cipher-images are uniformly distributed in $[0,255]$,
there are roughly $\lceil L/256 \rceil$ locations of the cipher-image pixels whose difference $(c_1(j)\dotminus c_2(j))$ equals $\Delta_p$},
i.e., each $v(i)$ has roughly $\lceil L/256 \rceil$ candidates.

Intuitively, more pairs of known plain-images help in eliminating {the} ambiguity of these candidates.
{To study this effect in a systematic way, we introduce the Self-Difference Matrix (SDM).
\begin{definition}[Self-Difference Matrix]
\label{def:sdm}
Given a sequence $\mathbf{P}_i = \{p_{k}(i)\}_{k=1}^{n}$,
the Self-Difference Matrix (SDM) of $\mathbf{P}_i$ is defined as follows:
\begin{equation}
\SDM(\mathbf{P}_i) =
\begin{pmatrix}
m_{1,1} &m_{1,2} & \dots  & m_{1,n} \\
m_{2,1} &m_{2,2} & \dots  & m_{2,n} \\
\vdots  &\vdots  & \ddots & \vdots \\
m_{n,1} &m_{n,2} & \dots  & m_{n, n} \\
\end{pmatrix},
\end{equation}
where
\begin{equation}
\nonumber
m_{r,c} =
\begin{cases}
(p_{r}(i) \dotminus p_{c}(i)), &\mbox{if } r < c; \\
0, &\mbox{if } r = c; \\
(p_{c}(i) \dotminus p_{r}(i)), &\mbox{if } r > c.
\end{cases}
\end{equation}
\end{definition}}

Suppose that there are $n$ pairs of known plain-images and the corresponding cipher-images,
which are denoted as $\mathbf{P} = \{P_k\}_{k=1}^{n}$ and $\mathbf{C} = \{C_k\}_{k=1}^{n}$,
respectively. According to the above analyses and Definition \ref{def:sdm},
we know that if $\SDM(\{p_{k}(i)\}_{k=1}^{n}) = \SDM(\{c_{k}(j)\}_{k=1}^{n})$,
$j$ is a possible solution of $v(i)$.

Initialize $i$ with $i=1$ and set $\mathbb{L} = [1, L]$, the procedures
of known-plaintext attack using $n$ pairs of known plain-images and the corresponding cipher-images can be described as follows:
\begin{description}[noitemsep, nolistsep]
\item[Step 1:] {Find $A_i =\SDM(\{p_{k}(i)\}_{k=1}^{n})$ using Definition~\ref{def:sdm}}.
\item[Step 2:] Find $\SDM(\{c_{k}(j)\}_{k=1}^{n})$ for all $j \in \mathbb{L}$. Determine the candidate set of $v(i)$ as
$\mathbb{S}=\{ j \in [1, L] \mid  \SDM(\{c_{k}(j)\}_{k=1}^{n}) =  A_i\}$, then randomly choose a candidate $j' \in \mathbb{S}$  and set $v(i)=j'$. Delete $j'$ from $\mathbb{L}$
to avoid conflict in the next round.
\item[Step 3:] If $i<L$, go to Step~1 and repeat the above operations.
\item[Step 4:] If $i=L$, compute $k(i)$ for all pixels by
\begin{equation}
\nonumber
k(i) = c(v(i)) \dotminus p(i).
\end{equation}
\end{description}
Once $\{k(i)\}_{i=1}^{L}$ and $\{v(i)\}_{i=1}^{L}$ are available, we can use them as the equivalent secret key to decipher any intercepted cipher-image encrypted with the same initial key.

The success of the above attack completely relies on Step~2, where we randomly choose a candidate from set $\mathbb{S}$.
We begin the theoretical analysis of the success rate with the following two trivial facts:
\begin{itemize}
\item The success rate rises as the number of known plain-images $n$ increases, i.e., the degree of freedom of SDM matrix, $ \tau = \frac {n\cdot (n-1)} {2}$, becomes larger.
\item If the cardinality of $\mathbb{S}$ satisfies $\# \{ \mathbb{S} \}=1$, it is confirmed that the $v(i)$ obtained is correct.
\end{itemize}
When $n=1$, the degree of freedom of SDM is $\frac {2\cdot (2-1)} {2} =1$. As explained before, there exists $\lceil L/256 \rceil$ candidates on the condition that pixels of the difference
image between the cipher-images are uniformly distributed in $[0,255]$.
For the special case $L=256$, i.e., the number of pixels is exactly equal to $256$, the uniformity of pixels of difference between the two cipher-images forces every integer in $[0,255]$ appear once and only
once\footnote{It should be noticed that pixels of the difference image between two cipher-images are not uniformly distributed in the encryption algorithm under study.
It is equal to the difference of the two corresponding plain-images, as pointed out in $\mathbf{Fact}$~\ref{fact2}.}.
Then in Step~2, we can find one and only one $j$ such that $\SDM(\{c_{k}(j)\}_{k=1}^{n}) =  A_i$ for certain $A_i$.
In other words, all $\{v(i)\}_{i=1}^{L}$ are derived accurately under this circumstance.

Let us consider the practical scenario that the degree of freedom of SDM satisfies $\tau  >1$ and the number of image pixels obeys $L>>256$.
As analyzed before, every valid entry of SDM has roughly $\lceil L/256 \rceil$ candidates. It is also noted that entries of SDM which have the same
gray value contribute nothing to further reduce $\# \{ \mathbb{S} \}$ in Step~2. Finally, based on the assumption that
pixels of difference image between cipher-images are uniformly distributed, we conclude that the attack will succeed with overwhelming probability if
\begin{equation}
\label{eq:success}
256^{\tau} \cdot \frac {(256)} {256} \cdot \frac {(256-1)} {256} \cdots  \frac {(256-(\tau-1))} {256} > L.
\end{equation}
For illustration purpose, we calculate the required number of known plain-images to cryptanalyze an intercepted cipher-image of size $512\times 512$.
In this case, $L = 512\times 512$. By Eq.~(\ref{eq:success}), one can easily find that $n \geq 3$ should be adopted.


Obviously, the computation complexity of the proposed attack is mainly caused by the iterations through Step~1 to Step~3.
To work out $A_i$ in Step~1, one needs to compute a symmetric SDM at the cost of $O(n^2)$. In Step~2, one needs to find $j'$ which satisfies
$\SDM(\{c_{k}(j')\}_{k=1}^{n}) =  A_i$. Then the rough computation complexity of Step~2 is $O(L)$. Step~3 needs the iteration of Step~1 and Step~2 for $L$ times.
Thus the overall complexity of this chosen-plaintext attack is $O(L^2\cdot n^2)$. As will be shown in the following simulations, $n=4$ is an empirical setting.
For images having a normal size, $L$ can reach $O(10^6)$. Thus the overall complexity of this algorithm could be as large as $O(10^{13})$, which is inefficient for practical
implementation. In the following discussion, we employ a simple strategy of trading space for time.
Instead of searching possible solutions for $v(i)$ one by one as described in Step~1 and Step~2, we pre-calculate all $\{A_i\}_{i=1}^{L}$ by $A_i=\SDM(\{p_{k}(i)\}_{k=1}^{n})$
and store the results as a sequence in the high-dimensional space. For each element of $\{A_i\}_{i=1}^{L}$, i.e., a SDM, we further map it to an integer between $1$ and $L$.
For any SDM's of the cipher-images, i.e., $\SDM(\{c_{k}(j)\}_{k=1}^{n})$, we perform the same mapping for this matrix to obtain an integer fall into the range $[1, L]$ and immediately turn to the corresponding
SDM of the plain-images who has the same mapping output. In this way, the computation complexity of the proposed attack is reduced from  $O(L^2\cdot n^2)$ to $O(L\cdot n^2)$ at the cost
of extra memory of size $O(L\cdot n^2)$.

\begin{figure}[!htb]
\centering
\subfigure[]{
    \label{fig:kpa:3dLenna}
    \begin{minipage}[t]{\imagewidth}
    \centering
    \includegraphics[width=\imagewidth]{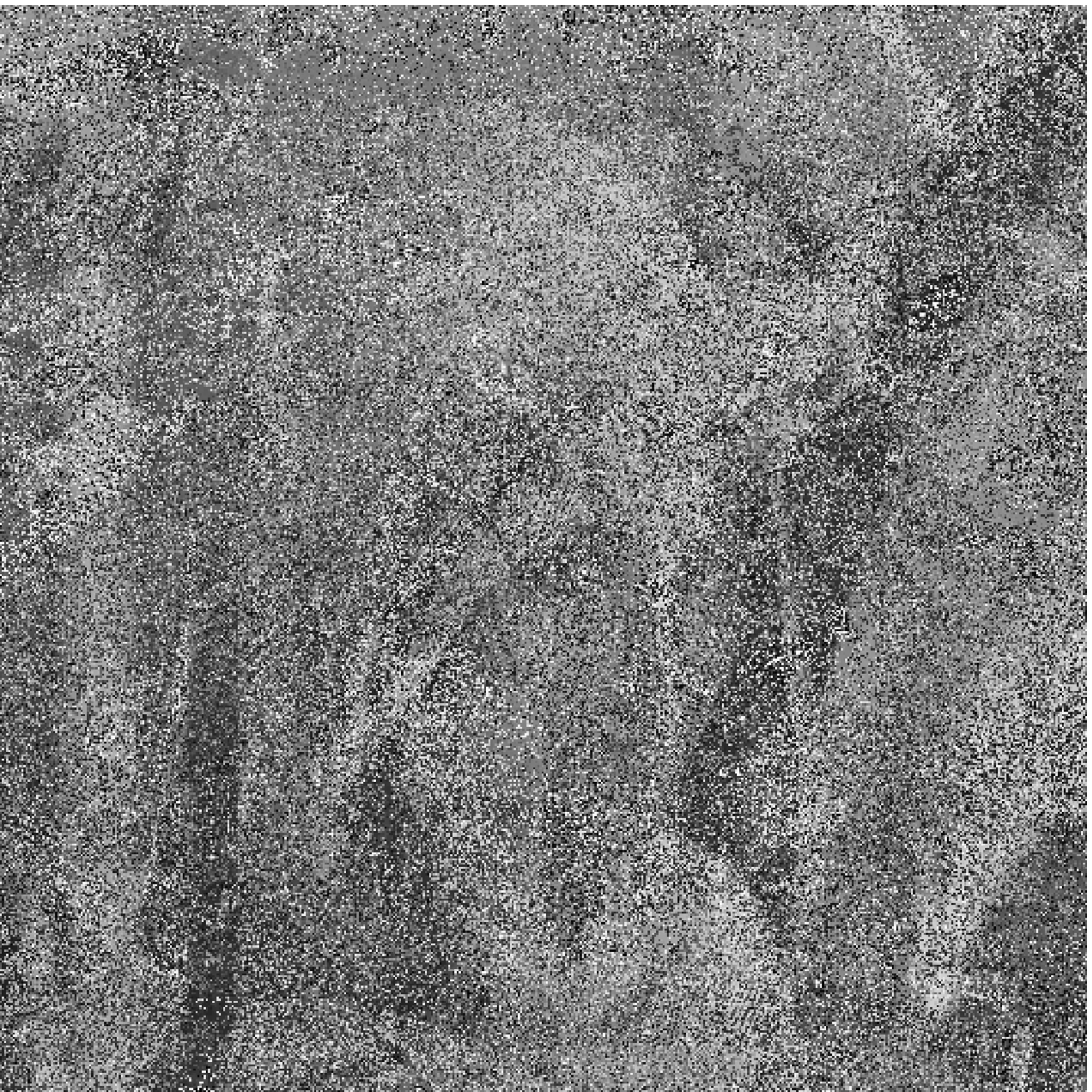}
    \end{minipage}}
\subfigure[]{
    \label{fig:kpa:4dLenna}
    \begin{minipage}[t]{\imagewidth}
    \centering
    \includegraphics[width=\imagewidth]{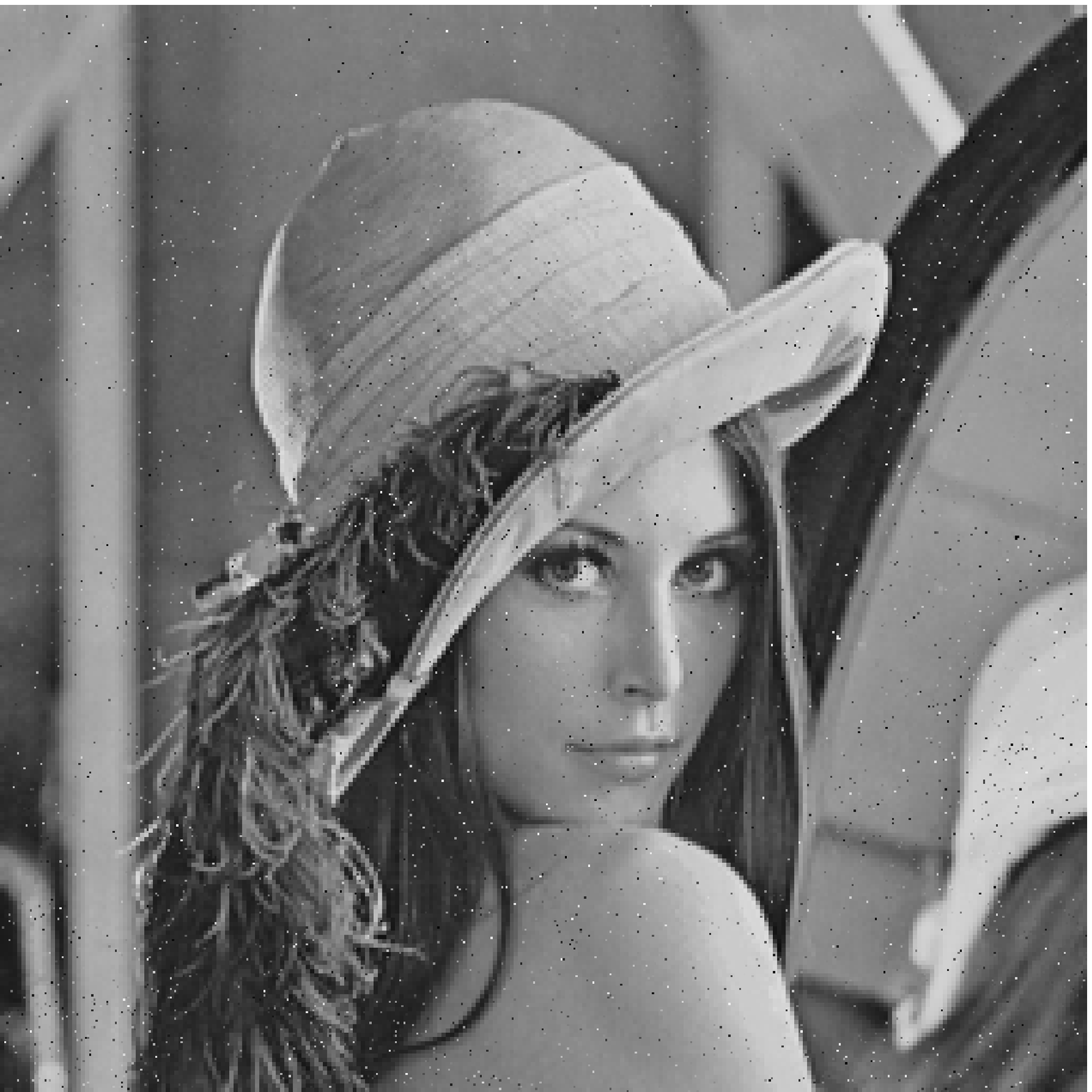}
    \end{minipage}}
\caption{The ``Lenna" image recovered from the cipher-image shown in Fig.~\ref{fig:cLenna} using the proposed known-plaintext attack with
(a) $3$ known plain-images;
(b) $4$ known plain-images.}
\label{fig:kpa}
\end{figure}

\begin{figure}[!htb]
\centering
\includegraphics[width=0.54\textwidth]{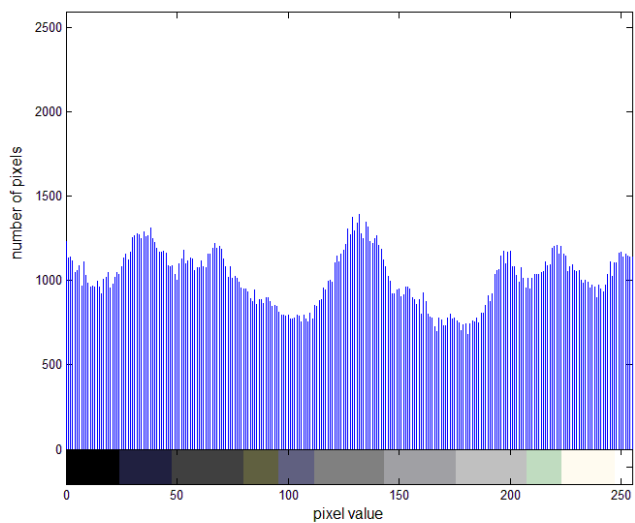}
\caption{Histogram of difference of two cipher-images corresponding to known plain-images ``Lenna" and ``Peppers". The key used is $(x_0, y_0, \mu_1, \mu_2, \gamma_1, \gamma_2) = (0.02145, 0.3678, 2.93, 3.17, 0.179, 0.139)$.}
\label{fig:histCipher}
\end{figure}

To verify the feasibility of the above known-plaintext attack, a lot of experiments have been carried out under the same key settings as employed in \cite[Sec.~4.1]{li2012image}.
The recovery results of Fig.~\ref{fig:cLenna} using $3$ and $4$ pairs of known plain-images are shown in Fig.~\ref{fig:kpa:3dLenna} and Fig.~\ref{fig:kpa:4dLenna},
respectively. Define the \textit{recovery rate} as
\begin{equation*}
\textit{recovery~rate} = \frac{number~of~correctly~recovered~pixels}{\textit{total~number~of~pixels}} \times 100\%,
\end{equation*}
and we found that the \textit{recovery rates} of Fig.~\ref{fig:kpa:3dLenna} and Fig.~\ref{fig:kpa:4dLenna} are $23.63\%$  and $98.45\%$, respectively.
It is clear that Fig.~\ref{fig:kpa:3dLenna} only contains a small amout of visual information of the original image, and we can barely figure out the contour of the original image.
However, the \textit{recovery~rate} reaches $98.45\%$ in Fig.~\ref{fig:kpa:4dLenna} and almost all subtle details can be observed.
The incorrectly recovered pixels can be treated as noise which can be eliminated by simple spatial filters. There are two reasons account for this mismatch between theoretical analyses and
experimental results: (1) Pixel distribution of the difference image between cipher-images corresponding to two known plain-images is not uniform, while our theoretical bound
are derived under the uniform distribution assumption. A typical example is shown in Fig.~\ref{fig:histCipher}.
(2) From Step~2 of the proposed attack, it can be found that a single incorrectly recovered $v(i)$ will double the error rate.

\begin{figure}[!htb]
\centering
\includegraphics[width=0.54\textwidth]{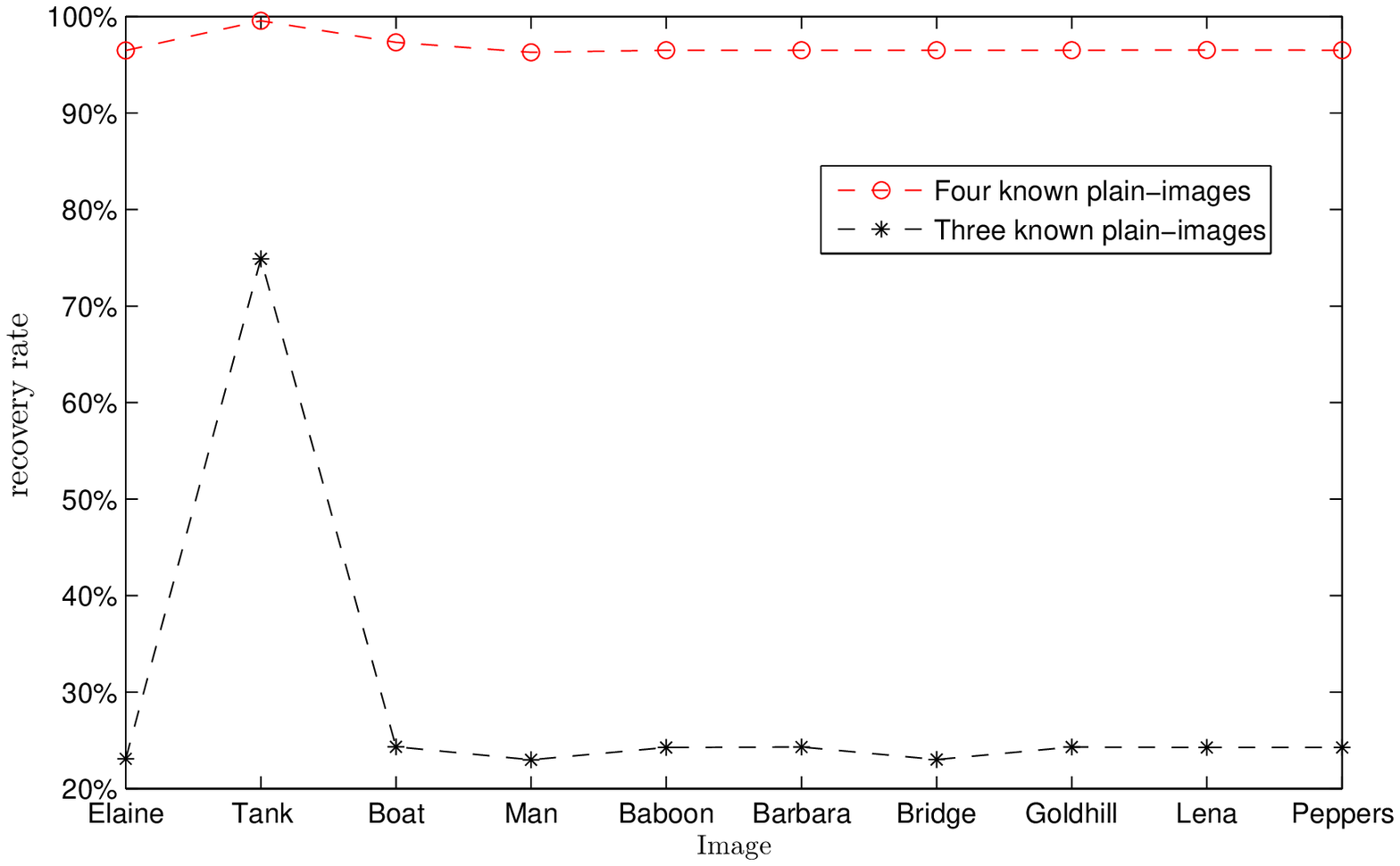}
\caption{The \textit{recovery rate} of the proposed known-plaintext attack using $3$ and $4$ known plain-images and the corresponding cipher-images.}
\label{fig:kpa2}
\end{figure}

To further study this phenomenon, more experiments were carried out on images having different textures using randomly generated secret keys.
The \textit{recovery rates} of $3$ and $4$ chosen-plain images are plotted in Fig.~\ref{fig:kpa2}.
It can be observed that the \textit{recovery rates} reach $95\%$ for all the test images when the number of known plain-images is $4$, while
the \textit{recovery rates} are around $25\%$ for almost all test images when the number of known plain-images is $3$. Thus, the extra known plain-image and its corresponding cipher-image
can be considered as a penalty term to bridge the gap between theoretical analysis and practical implementation.

\section{Conclusion}
The complexity for breaking an image cipher based on scrambling and Vegin{\`e}re cipher has been analyzed.
In the chosen-plaintext attack scenario, we propose the optimal chosen plain-image attack by improving the previous work suggested by Zhang \textit{et al}.
In the known-plaintext attack scenario, we present an efficient known plain-image attack which makes use of the so called self-difference matrix. The required number of known-images
to guarantee a successful attack has been worked out theoretically. Some practical considerations of this attack are also described for the purpose of
implementing it on a personal computer.

\begin{acknowledgements}
This work was supported by the National Natural Science Foundation of China (Nos. 60673193 and 60083001).
\end{acknowledgements}

\bibliographystyle{spmpsci}
\bibliography{MTA}
\end{document}